\documentclass[twocolumn,showpacs,prb,floatfix]{revtex4}
\usepackage{epsfig}

\newcommand{\be}{\begin{equation}}
\newcommand{\ee}{\end{equation}}
\newcommand{\beqn}{\begin{eqnarray}}
\newcommand{\eeqn}{\end{eqnarray}}

\newcommand{\chisg}{\chi_{_{\mathrm{SG}}}}

\begin{document}

\title{On the existence of a finite-temperature transition in the 
two-dimensional gauge glass}
\author{Helmut G.~Katzgraber}
\affiliation{Theoretische Physik, ETH H\"onggerberg,
CH-8093 Z\"urich, Switzerland}

\date{\today}

\begin{abstract}
Results from Monte Carlo simulations of the two-dimensional gauge
glass supporting a zero-temperature transition are presented. A
finite-size scaling analysis of the correlation length shows that
the system does not exhibit spin-glass order at finite temperatures.
These results are compared to earlier claims of a finite-temperature
transition.
\end{abstract}

\maketitle

The existence of a finite-temperature spin-glass transition in the
two-dimensional gauge glass has been a source of great controversy.  
While the existence of a transition in three dimensions is
undisputed,\cite{olson:00} several different approaches\cite{fisher:91,reger:93,hyman:95,kosterlitz:97,granato:98,choi:99,kim:00,katzgraber:02a,holme:03}
have been made to determine the spin-glass ordering temperature 
$T_c$ in two dimensions with different results. As the gauge glass 
in two dimensions is a model commonly used to describe the 
vortex-glass transition in dirty high-temperature superconductor 
thin films,\cite{blatter:94} it is important to settle the 
question of whether the system exhibits spin-glass order 
at finite temperatures or not. Experimental results
on YBCO thin films\cite{dekker:92} show no vortex-glass order,
whereas bulk materials do exhibit a transition to a glassy phase.  
A brief summary of the different approaches used to determine $T_c$
is presented, followed by results from a novel
technique\cite{ballesteros:00} that strongly supports the absence
of a glassy phase at finite temperatures in the two-dimensional
gauge glass.

Early work by Fisher {\em et al.}\cite{fisher:91}~who study the
size-dependence of domain-wall energies with a zero-temperature 
domain-wall renormalization group method, shows evidence of a 
zero-temperature transition in the two-dimensional gauge glass. 
Gingras,\cite{gingras:92} as well as Akino and Kosterlitz,\cite{akino:02} 
who also study the stiffness exponent of the model, find agreeing
results. 

In the aforementioned work,\cite{fisher:91} Fisher {\em et al}.~also 
analyze the scaling of Binder ratios\cite{binder:81} of the order 
parameter and obtain further evidence of a zero-temperature transition.
These results are in agreement with Monte Carlo simulations by Reger 
and Young\cite{reger:93} who conclude that the transition temperature is 
not finite by studying the scaling of the currents induced by a twist in 
the boundary conditions. Katzgraber and Young\cite{katzgraber:02a}
find evidence of a zero-temperature transition for the
two-dimensional gauge-glass Hamiltonian with periodic boundary
conditions by studying the scaling of the currents and the
spin-glass susceptibility. They also conclude that one can scale
data for the susceptibility well for any positive $T_c$, probably
due to the extra fitting parameter $\eta$, the anomalous dimension
of the spin-glass order parameter $q$, and show that studying the
scaling of the spin-glass susceptibility alone is not enough to
obtain evidence of a finite-temperature transition. This may explain
why similar simulations by Choi and Park,\cite{choi:99} who study
the susceptibility only, find a finite-temperature transition with
$T_c = 0.22 \pm 0.02$.

Furthermore, Kim\cite{kim:00} has studied the gauge glass with
fluctuating twist boundary conditions\cite{olsson:92} by means of
resistively shunted junction (RSJ) dynamics and finds a finite $T_c$
in agreement with the work of Choi and Park,\cite{choi:99} and
with early results by Li\cite{Li:92} (challenged later by 
Simkin\cite{simkin:96}). Surprisingly, Granato,\cite{granato:98} 
who also uses RSJ dynamics
but with periodic boundary conditions, obtains $T_c = 0$.

More recently Holme {\em et al.}\cite{holme:03} have studied the
two-dimensional gauge glass by means of Monte Carlo simulations with
fluctuating twist boundary conditions. By introducing an extra
anomalous exponent $b$ they are able to scale data for the currents
and the helicity modulus and so obtain a finite transition
temperature close to $0.20$.

As there are diverging opinions on the spin-glass transition
temperature of the two-dimensional gauge glass, the issue is
reconsidered in the present work using a different approach. The
scaling of the finite-system correlation lengths is studied, giving
compelling evidence that $T_c \approx 0$, at least for the system
sizes studied here. These results are supported by data for the
spin-glass susceptibility.

The gauge-glass Hamiltonian,\cite{shih:84} which describes a
disordered granular superconductor, is given by
\begin{equation}
{\cal H} = -J \sum_{\langle i, j\rangle} \cos(\phi_i - \phi_j - A_{ij}),
\label{hamiltonian}
\end{equation}
where the sum ranges over nearest neighbors on a square lattice of
size $N = L^2$ and $\phi_i$ represent the angles of the $XY$ spins.
The $A_{ij}$ are quenched random variables uniformly distributed
between $[0,2\pi]$ with the constraint that $A_{ij} = - A_{ji}$. In
this work $J = 1$ and periodic boundary conditions are applied.

Previously\cite{katzgraber:02a} we estimated the critical
temperature of the two-dimensional gauge glass by studying the
scaling of the currents\cite{reger:91} induced by a twist in the
boundary conditions. This has an advantage over studying the
crossing of Binder ratios\cite{binder:81} because the latter will
not splay enough for $XY$ systems\cite{olson:00} at $T < T_c$,
making it almost impossible to accurately determine the crossing
point. Studying the currents requires some care as well, as these
have different finite-size scaling forms\cite{katzgraber:02a}
depending on a finite or zero transition temperature.  Finally, the
currents show strong statistical fluctuations and so require a large
amount of disorder realizations. Therefore in this work the scaling
of the correlation lengths is studied as these show little
statistical fluctuations, have a simple finite-size scaling form and
splay out well for $T < T_c$.
\begin{figure}
\centerline{\epsfxsize=\columnwidth \epsfbox{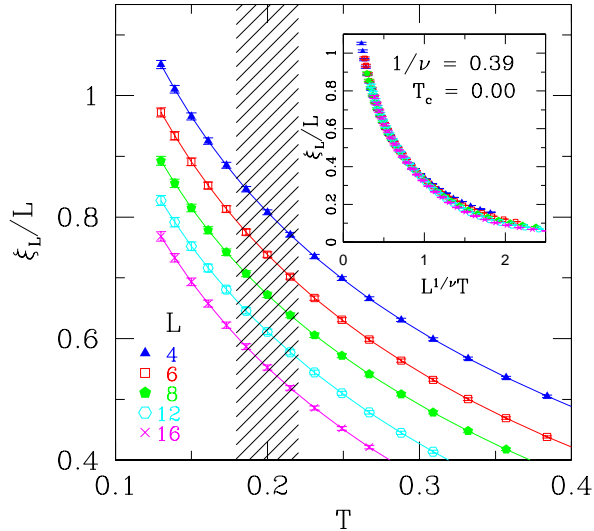}}
\vspace{-1.0cm}
\caption{
Correlation length $\xi_L$ divided by $L$ for different system
sizes. At all temperatures the data decrease with increasing $L$.
This indicates that if $T_c$ is finite, it must be less than $0.13$,
the lowest temperature simulated. The shaded region represents the
area around $T \approx 0.20$ where the
claimed (Refs.~\onlinecite{holme:03,choi:99,kim:00}) 
finite-temperature transition takes place. 
The inset shows a scaling plot of the data for
$\xi_L/L$ plotted against $L^{1/\nu}T$ for $T_c = 0$ and $1/\nu =
0.39$. One can see that the data scale very well, especially for low
temperatures.
}
\label{xiLplot}
\end{figure}
This method has been applied with success to Ising,\cite{ballesteros:00} 
as well as vector spin glasses.\cite{lee:03} The spin-glass order 
parameter for the gauge glass is given by
\begin{equation}
q = \frac{1}{N} \sum_j^N \exp[i(\phi_j^{\alpha} - \phi_j^{\beta})] \; ,
\label{qdef}
\end{equation}
where ``$\alpha$'' and ``$\beta$'' represent two replicas of the
system with the {\em same} disorder. Equation (\ref{qdef}) is
generalized to finite wave vectors ${\bf k}$:
\begin{equation}
q({\bf {k}}) = \frac{1}{N} \sum_j^N \exp[i(\phi_j^{\alpha} - 
\phi_j^{\beta}) + i {\bf k} \cdot {\bf R}_j] \; .
\label{qkdef}
\end{equation}
The wave vector dependent spin-glass susceptibility $\chisg({\bf
k})$ is then given by
\begin{equation}
\chisg({\bf k}) = N[\langle |q({\bf k})|^2\rangle]_{\rm av} \; .
\label{chisg}
\end{equation}
Here, $\langle \cdots \rangle$ denotes a thermal and $[\cdots]_{\rm
av}$ a disorder average. The finite-system correlation length $\xi_L$
is determined from an expansion\cite{cooper:82,ballesteros:00,comment:2}
of the spin-glass susceptibility for small ${\bf k}\xi_L$ in the framework of
an Ornstein-Zernicke approximation\cite{martin:02}
\begin{equation}
[\chisg({\bf k})/\chisg({\bf 0})]^{-1} = 1 + ({\bf k}\xi_L)^2 + O[({\bf k}\xi_L)^4] \; .
\label{OZ}
\end{equation}
Keeping only the leading term in the expansion and taking into account 
the lattice periodicity as well as corrections to scaling, we obtain
\begin{equation}
\xi_L = \frac{1}{2\sin(|{\bf k}_{\rm min}|/2)}
\left[ \frac{\chisg({\bf 0})}{\chisg({\bf k}_{\rm min})} 
- 1\right]^{1/2} \; ,
\label{xiL}
\end{equation}
where ${\bf k}_{\rm min} = (2 \pi/L,0,0)$ is the smallest nonzero
wave vector. In Eq.~(\ref{xiL}), $\chisg({\bf 0})$ is the standard
spin-glass susceptibility. 

The finite-size scaling of $\xi_L$ can be understood by studying
the finite-size scaling of the spin-glass susceptibility
extended to finite wave vectors $k$
\begin{equation}
\chisg(T,L,k) = L^{2-\eta}\tilde{C}[L^{1/\nu}(T - T_c),kL] \;.
\label{chiscale1}
\end{equation}
From Eq.~(\ref{OZ})
$ k_{\rm min}^2\xi_L^2 \sim \chisg(0)/\chisg(k_{\rm min})$
up to an additive constant, and so the $L$-dependent prefactor 
in Eq.~(\ref{chiscale1}) 
cancels. The dependence of $\chisg$ on $kL$ drops out as well, 
as in Eq.~(\ref{xiL}) $k = k_{\rm min} = 2\pi/L$. We therefore 
obtain the simple finite-size scaling form
\begin{equation}
\xi_L/L = {\tilde X}[L^{1/\nu}(T - T_c)] \; ,
\label{xiscale}
\end{equation}
and so, the introduction of an anomalous exponent\cite{holme:03} is
not plausible. This has also the advantage that the finite-size
scaling has few parameters.  From Eq.~(\ref{xiscale}) we see that
data for different system sizes $L$ will cross at $T = T_c$. If $T_c
= 0$, one expects that data for different $L$ will decrease with
increasing $L$ and not cross for any finite temperature.

For the simulations the parallel tempering Monte Carlo
method\cite{hukushima:96,marinari:98b} is used. Because the
equilibration test for short-range spin glasses introduced by
Katzgraber {\em et al.}\cite{katzgraber:01,katzgraber:01c,lee:03}
does not work for the gauge glass\cite{katzgraber:01c} as the
disorder is not Gaussian, equilibration is tested by the traditional
technique of requiring that different observables are independent of
the number of Monte Carlo steps. By doubling the number of Monte
Carlo sweeps between each measurement until the last three agree
within error bars, equilibration is ensured.  The equilibration
times used for the different system sizes are listed in
Ref.~\onlinecite{katzgraber:02a}, Table I. For the largest size
studied, $L_{\rm max} = 16$, $10^3$ disorder realizations are
performed, whereas for the smaller sizes $L$, $10^3 (L_{\rm max}/L)$
realizations. The lowest temperature studied is $T = 0.13$, well
below the claimed\cite{holme:03,choi:99,kim:00} transition at $T
\approx 0.20$. Here, 30 temperature replicas are used with the
highest temperature being $T = 1.058$.

Figure \ref{xiLplot} shows data for $\xi_L/L$ for different system
sizes $L$. The data decrease with increasing $L$ for all
temperatures. In particular, there is {\em no} crossing of the data
around $T \approx 0.20$ (shaded area) where the spin-glass
transition is supposed to take place.\cite{holme:03,choi:99,kim:00}
The inset of Fig.~\ref{xiLplot} shows a finite-size scaling plot of
the data according to Eq.~(\ref{xiscale}). One can see that the data
scale very well for $T_c = 0$ and $1/\nu = 0.39 \pm 0.02$, values
found in Ref.~\onlinecite{katzgraber:02a} by scaling the currents
and the spin-glass susceptibility. The error bar in $1/\nu$ is
determined by varying its value until the data do not scale well. An
attempt to scale the data in Fig.~\ref{xiLplot} to any value of a
finite $T_c$, in particular $0.20$, fails for any choice of $\nu$.

\begin{figure}
\centerline{\epsfxsize=\columnwidth \epsfbox{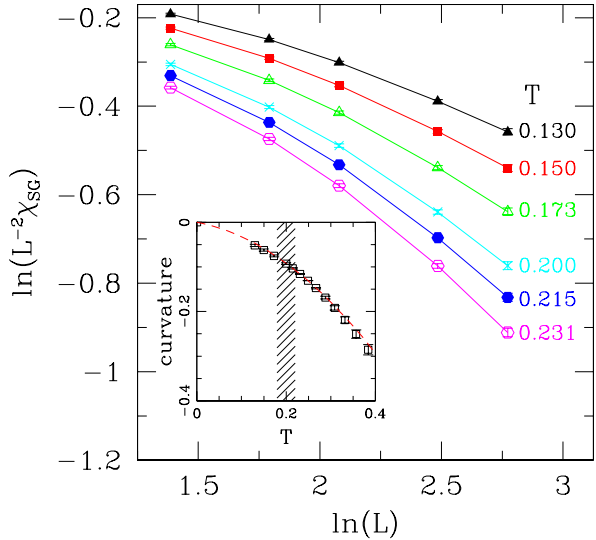}}
\vspace{-1.0cm}
\caption{
Data for $\ln(\chisg/L^2)$ vs $\ln(L)$ for different temperatures.
At $T_c$ one expects $\chisg \sim L^{2-\eta}$. The data show a
strong downward curvature indicating that $T > T_c$ for the range of
temperatures shown. This is emphasized in the inset where the
curvature (from a second-order polynomial fit) is displayed. At $T_c$
one expects the curvature to cross zero. Here the data extrapolate
to $T_c \approx 0$ (the dashed line represents a second-order 
polynomial fit extrapolation).
The shaded area shows the region where one would expect the curvature to
cross zero if $T_c \approx 0.2$.
A finite-size scaling analysis of the spin-glass
susceptibility according to Eq.~(\ref{chiscale}) works well for low
enough temperatures with $T_c = 0$, $\eta = 0$, and $1/\nu = 0.39$,
in agreement with results from Ref.~\onlinecite{katzgraber:02a}.
}
\label{logchi}
\end{figure}

Standard finite-size scaling predicts that the spin-glass
susceptibility scales as\cite{comment:1,carter:03}
\begin{equation}
\chisg = L^{2-\eta}\tilde{C}[L^{1/\nu}(T - T_c)] \;,
\label{chiscale}
\end{equation}
which means that at $T = T_c$ it should scale as a power law $\sim
L^{2-\eta}$.  In Fig.~\ref{logchi} data for $\ln(\chisg/L^2)$
vs $\ln(L)$ is shown.  If the data were at $T_c$, one would expect
a straight line. In particular, $T = 0.20$ shows a strong downward
curvature indicating that the system is {\em above} the critical
transition temperature. A fit of the data in Fig.~\ref{logchi} to a
straight line for the different temperatures yields quality of fit
probabilities\cite{press:95} between $10^{-9}$ for the lowest, and
$10^{-20}$ for the highest temperature, respectively. A second-order
polynomial fit to the data, instead, has quality of fit
probabilities $\sim 0.9$. The curvature of the second-order fits is
determined and shown in the inset of Fig.~\ref{logchi}. Because at
$T < T_c$ [$T > T_c$] one expects $\chisg(L)$ to bend upward
[downward], the curvature has to go through zero at $T_c$. The data
in the inset of Fig.~\ref{logchi} do not cross the horizontal axis
for the temperature range studied and extrapolate within error bars
to $T_c \approx 0$. This is in agreement with results by Katzgraber
and Young\cite{katzgraber:02a} who obtain good scaling of the
susceptibility at low temperatures for $T_c = 0$, $\eta =0$, and
$1/\nu = 0.39$.

To conclude, evidence of a zero-temperature transition in the
two-dimensional gauge glass with periodic boundary conditions is
presented by studying the finite-size scaling of the correlation
lengths. This approach is better than previous methods used because
the correlation lengths show small statistical fluctuations and have
a simple finite-size scaling form. In addition, the curvature of the 
spin-glass susceptibility as a function of system size indicates that 
there is no spin-glass transition at $T \approx 0.20$, at least for 
the system sizes studied. Moreover the curvature extrapolates within 
error bars to $T_c \approx 0$. These results are in agreement with 
experimental data on YBCO thin films,\cite{dekker:92} where a 
finite-$T$ is absent, and different numerical
approaches.\cite{fisher:91,reger:93,akino:02,katzgraber:02a}  In
particular, it is shown that the system does {\em not} exhibit a
spin-glass transition at $T \approx 0.20$, as claimed in
Refs.~\onlinecite{choi:99}, \onlinecite{kim:00}, and
\onlinecite{holme:03}.  It is noteworthy that
Refs.~\onlinecite{kim:00} and \onlinecite{holme:03} find the
aforementioned finite temperature transition by studying the gauge
glass with fluctuating twist boundary conditions instead of
the commonly used periodic boundary conditions. Could this be one source
of the discrepancies?

Finite-size effects can influence the data, and so an analysis with
larger system sizes is desirable. Nevertheless, the data presented
here strongly support the absence of a transition at $T \approx
0.20$, although a transition at $T < 0.13$, the lowest temperature
studied, while unlikely the case, cannot be ruled out
completely.\newline

I would like to thank G.~Blatter, I.~Campbell, M.~Troyer, and
A.~P.~Young for helpful discussions and to K.~Tran for carefully
reading the manuscript. The simulations were performed on the 
Asgard cluster at ETH Z\"urich.

\bibliography{refs,comments}

\end{document}